# An Approach for Normalizing Fuzzy Relational Databases Based on Join Dependency

## Deepa S.

*Assistant Prof. , DoS in Computer Application,  Pooja Bhagavat Memorial Mahajana PG Centre, KRS Road, Metagalli, Mysore, Karnataka, India.*

## ABSTRACT

*Fuzziness in databases is used to denote uncertain or incomplete data. Relational Databases stress on the nature of the data to be certain. This certainty based data is used as the basis of the normalization approach designed for traditional relational databases. But real world data may not always be certain, thereby making it necessary to design an approach for normalization that deals with fuzzy data. This paper focuses on the approach for designing the fifth normal form (5NF) based on join dependencies for fuzzy data. The basis of join dependency for fuzzy relational databases is derived from the basic relational database concepts. As join dependency implies an multivalued dependency by symmetry the proof of join dependency based normalization is stated from the perspective of multivalued dependency based normalization on fuzzy relational databases.*

## 1. Introduction

Normalization approach aims to generate a set of relation schemas that allows us to store information without necessary redundancy. Relational model usually take care of only well defined data. However real world data is often ambiguous in nature since they are closer to human intuitions. In order to capture more meaning to the data an extension of the classical relational model called the fuzzy relational model was proposed. Normalizing fuzzy data needs to examine the nature of fuzziness[4]. Fuzziness of data may range from incomplete to uncertain form of data. Examining the nature of fuzziness needs to analyze the possibility of functional dependency and Multivalued Dependency(MVD) existent in the relation schema. Several studies are made on relational model from the perspective of fuzzy set theory by extending the relational algebra concepts to suit the fuzzy databases. This paper examines fuzzy relations based on similarity measures. The data set examined is basically multivalued in nature. In view of this the paper is organized as follows: Section 2 of the paper deals with the concepts of fuzziness on relations. Section 3 defines join dependency on fuzzy relational schema. In section 4 the result analyzes the proposed join dependency on fuzzy relational schema and states the Fifth normal form.

## 2. Fuzzy relations

An n-ary fuzzy relation r is a fuzzy subset of the Cartesian product of some universe. A fuzzy relation is defined as the Cartesian product of the list of domains.$D_1, D_2, D_3,\text{----} D_n$. Let S= $A_1, A_2, A_3,\text{----} A_n$ be attributes concerned to these domains, then an operator that tries to deal with the fuzziness on data can be applied to retrieve multiple or ranged values.

Fuzziness on data is applied by generalizing the values assigned to the elements of the universal set to fall within a specified range. The larger the values the higher the degree of membership. Such a function that assigns the values to the elements of the universal set is called a membership function and the set is called a fuzzy set. The general notion for representing the fuzzy set is [1]:

$$\mu A : X [0, 1] \quad (1)$$

although the set [0,1] is totally ordered , sets $[0,1]^n$ for any $n \geq 2$ are ordered only partially.

## 3. Join dependency on relational  model

Join dependency on relational model is a special case of multi valued dependency[5]. The same idea is extended to Fuzzy relational databases[2]. That is a Join Dependency(JD) denoted by JD(R1,R2) implies

an:
$MVD(R1 \cap R2) \rightarrow\rightarrow$
$(R1 - R2) \text{ or } (R2 - R1))$.  (2)

3.1 Fuzzy Relational Join Dependency

Therefore the concepts of MVD involving multivalued data with X and Y as single or set of attributes belonging to R in symbolic Relational Databases as proposed in Deepa S(2011):

If for all tuples t1 and t2,

$the\ min\ (ds\ (t1[x], t2[x]))$
$\leq min\ (ds\ (t1[Y], t2[Y])), then\ min\ (ds\ (t1[x], t2[x]))$
$\leq (ds\ (R$
$- min(ds\ (t1[X], t2[X])), min\ (ds\ (t1[Y], t2[Y])))$

Where ds is the degree of similarity that exists between the tuples. The following similarity measure for multivalued data has been proposed in Nagalakshmi and Suhasini (2010) [3]:

$If\ \{x \cap Y\} = \emptyset\ sim\ (X, Y) = 0$

Else

$Sim(X \rightarrow Y) = 1 - \left\{ \dfrac{|\{X\} - \{X \cap Y\}|}{|\{X \cup Y\} - \{X \cap Y\}|} \right\}$

$Sim(Y \rightarrow X) = 1 - \left\{ \dfrac{|\{Y\} - \{X \cap Y\}|}{|\{X \cup Y\} - \{X \cap Y\}|} \right\}$

Extending the concepts of JD to the fuzzy relational database environment the MVD can be stated as:
$MVD(R1 \cap R2) \sim \rightarrow\rightarrow (R1 - R2)$.  (3)

If for all tuples t1 and t2 in R1 and t3 and t4 in R2 the join dependency for fuzzy relational databases can be defined
$min(ds\ (t1[x], t2[x])) \leq$
$ds\ (R - min(ds\ (t1[X], t2[X])),$

$min\ (ds\ (t1[Y], t2[Y])$
$\cap\ min\ (ds\ (t3[X], t4[X]))$
$\leq ds\ (R - min(ds(t3[X], t4[X])),$

$min\ (ds\ (t3[Y], t4[Y])$.  (4)

## 4. Discussion of results

Consider a fuzzy relation schema of the form Supply(Supplier_name, part_name, project_name) that provides information regarding the supply of parts concerned with particular projects. Here domain(supplier_name) is an ordinary set and domain(part_name, project_name) are fuzzy sets.

### Table 1: An instance of supply

| Supplier_name | Part_name | Project_name |
|---|---|---|
| ABC | P1,P2 | Proj X, Proj Y |
| MNO | P1,P3 | Proj X, Proj Y |
| XYZ | P2 | Proj Z |

The above schema specifies a join dependency of three relations (Supplier_name, Part_name) , (Supplier_name, Project_name) and (Part_nme, Project_name).

The ds for the fuzzy attributes can be derived from (2) and (3) and the degree of similarity of part_name and project_name in instance supply is shown below:

### Table 2: Degree of similarity of the multivalued attribute part_name

|  | P1,P2 | P1,P3 | P2 |
|---|---|---|---|
| P1,P2 | 1 | 0.34 | 0 |
| P1.P3 | 0.34 | 1 | 0.34 |
| P2 | 0 | 0.5 | 1 |

**Table 3: Degree of similarity of the multivalued attribute part_name**

|  | Proj X, Proj Y | Proj X, Proj Y | Proj Z |
|---|---|---|---|
| Proj X, Proj Y | 1 | 1 | 0.33 |
| Proj X.Proj Y | 1 | 1 | 0.33 |
| Proj Z | 0.7 | 0.7 | 1 |

Comparing the ds based on the join dependency derived in (4) on the three possible projections R1(supplier_name,part_name), R2(supplier_name,project_name) and R3(part_name, project_name) of the supplier schema leads to the following results:

ds for R1 and R2 will give the result based on $MVD\ (R1\ \cap\ R2) \sim \rightarrow\rightarrow (R1 - R2)$ as

$min(0), min(1)$

$\leq ds\ (R - min(1,1)), ds(R - 1)$

Similarly ds for R2 and R3 will give the result based on $MVD\ (R2\ \cap\ R3) \sim \rightarrow\rightarrow (R2 - R3)\ as$

$min(0), min(0.7)$

$\leq ds\ (R - min(1,0.33)), ds(R - 0.7)$

similarly ds for R3 and R1 will give the result based on $MVD\ (R3\ \cap\ R1) \sim \rightarrow\rightarrow (R3 - R1)\ as$

$min(0.34), min(0.33) \leq$

$ds(R- in(1,0.7)), ds(R - 0.3)$, there by proving that the join dependency implies an multivalued dependency by symmetry. The ds based comparisons proves that the minimum of the ds of the compared tuples will always show a similarity value less than or equal to similarity values in the tuples of the corresponding projections[6].

Based on the results the fifth normal form for the fuzzy relational databases can be stated as :

A relation schema R is said to be in fifth normal form with respect to a set of functional, multivalued and join dependencies if there exists a trivial join dependency where the determinant of the join dependency becomes the superkey of the relation R.[8]

## 5. Conclusion

This paper deals with the design of join dependency based normalization as a broad conceptual framework for dealing with uncertainty of information. Dealing with uncertain data is one of the challenging task in handling real world data. The form of uncertainty dealt in the paper is limited to multivalued data. Though uncertainty may range from qualitative to interval based values, the possibility of deriving fuzzy data based normalization from the other perspective of uncertainty forms the future scope of work.